\documentclass[12pt,a4paper]{article}
\usepackage{jheppub}
\usepackage{dsfont}
\usepackage{amssymb,amsmath}
\usepackage{epsfig}
\usepackage{epstopdf}
\usepackage{latexsym}
\usepackage{graphicx}
\usepackage{subfig}
\usepackage{booktabs}
\usepackage{bbm}
\usepackage{color}
\pdfoutput=1
\makeatletter
\def\@fpheader{\relax}
\makeatother

\def\be{\begin{equation}}
\def\ee{\end{equation}}

\title{Scale invariance with fundamental matters and anomaly: A holographic description}
\author{Avik Banerjee$^a$, Arnab Kundu$^{a}$, Augniva Ray$^a$}
\affiliation{$^{a}$Theory Division, Saha Institute of Nuclear Physics, HBNI, 1/AF Bidhannagar, Kolkata 700064, India.}
\emailAdd{avik.banerjee[at]saha.ac.in, arnab.kundu [at] saha.ac.in, augniva.ray[at]saha.ac.in}

\abstract{Generally, quantum field theories can be thought as deformations away from conformal field theories. In this article, with a simple {\it bottom up} model assumed to possess a holographic description, we study a putative large $N$ quantum field theory with large and arbitrary number of adjoint and fundamental degrees of freedom and a non-vanishing chiral anomaly, in the presence of an external magnetic field and with a non-vanishing density. Motivated by the richness of quantum chromodynamics under similar condition, we explore the solution space to find an infinite class of scale-invariant, but not conformal, field theories that may play a pivotal role in defining the corresponding physics. In particular, we find two classes of geometries: Schr\"{o}dinger isometric and warped AdS$_3$ geometries with an ${\rm SL}(2,R) \times {\rm U}(1)$ isometry. We find hints of spontaneous breaking of translational symmetry, at low temperatures, around the warped backgrounds.}

\begin{document}

\maketitle
\flushbottom

\section{Introduction \& conclusions}

Quantum Field Theory (QFT) is best understood as the framework of describing a class of phenomena at a given energy-scale, with a particular model. Quantum Chromodynamics (QCD) is such a model that describes the strong interaction between quarks and gluons. Despite its enormous success, QCD remains a challenge till date, simply because the coupling is large at low energies. At the same time, the infrared (IR) physics of QCD is quite interesting, both empirically and based on intriguing theoretical possibilities. Much of such physics is outside the scope of conventional perturbative QFT techniques and one may adopt {\it effective theories} or lattice simulations to address various key aspects of the dynamics of quarks and gluons.

Of particular interest, for example, is the existence of a colour-flavour locked phase of QCD, which is commonly known as the colour superconductivity\cite{Alford:1997zt}, which can be characterized by a symmetry breaking: ${\rm SU} (3)_{\rm c} \times {\rm SU}(3)_{\rm L} \times {\rm SU}(3)_{\rm R} \to {\rm SU}(3)_{\rm c+L+R}$. Here ${\rm SU}(3)_{\rm c}$ corresponds to the colour gauge group, ${\rm SU}(3)_{\rm L/R}$ corresponds to the global flavour symmetry group. As a result, the emergent physics is invariant under a simultaneous colour and flavour rotation, and one requires the number of these two degrees of freedom at equal strength for this mixing to take place.

In this article, we will adopt holography or gauge-gravity duality as a framework of studying strongly coupled QFT, specially SU$(N_c)$-type gauge theories with matter in the adjoint and the fundamental sectors, in the limit where the number of the adjoint and the fundamental degrees of freedom are of the same order. We will, however, not attempt to provide the precise UV-complete description of the model under consideration, in terms of strings and branes degrees of freedom. Instead, we intend to capture one particular aspect with minimal and hopefully sufficiently generic ingredients.

Towards that, we will study a simple generalization of the model already discussed in \cite{Kundu:2016oxg}, see {\it e.g.}~\cite{Pal:2012zn, Tarrio:2013tta} for earlier works, by including a non-vanishing Chern-Simons term in $(4+1)$-bulk dimensions. Thus, the model in consideration consists of Einstein-Hilbert action with a negative cosmological term, a Dirac-Born-Infeld (DBI) matter field and a non-vanishing Chern-Simons term. In the putative dual field theory, this is expected to capture the dynamics of $N_c$ adjoint and $N_f$ fundamental matter fields, for a fixed value of $\left( N_f/ N_c \right)$, along with the presence of an anomaly. This anomaly is reminiscent of the triangle anomaly in QCD, based on the global transformation properties of the corresponding Chern-Simons term with respect to discrete symmetries, such as parity or time-reversal. See \cite{Son:2009tf} for an early discussion on the effect of triangle anomaly on hydrodynamic properties of a strongly coupled gauge theory, and \cite{Erdmenger:2008rm, Banerjee:2008th} for an AdS/CFT derivation of the transport coefficients induced by anomalies. We consider precisely the same term as in the studies above, as well as in \cite{D'Hoker:2009bc}.

Before proceeding further, a few comments are in order: Note that, we will only consider the Abelian version of the DBI action. This is primarily for convenience, however, upon including non-Abelian features, the physics may qualitatively change. Thus, we effectively describe an ${\rm U}(1)^{N_f}$ symmetric configuration in the fundamental sector, as opposed to an ${\rm U}(N_f)$ symmetric one. Furthermore, the dimensionality of the bulk (Einstein-Hilbert) gravity action and the DBI action are taken to be the same. This is suggestive that, in terms of a potential D-brane picture, one is considering a space-filling D-brane matter to capture the dynamics of the fundamental sector. Such a configuration may be obtainable starting from a stack of $N_c$ D$3$-branes and $N_f$ D$9$-branes (or anti-branes). Intuitively, considering a D$3$-D$9$ brane system, with such a perspective, is a natural choice. The decoupling limit of the D$3$ branes provides us with the pure-glue sector of the theory, and any D-brane configuration that one is expected to eventually end up with in {\it e.g.}~type IIB supergravity can be obtained from a D$9$-$\overline{{\rm D}9}$ system by virtue of tachyon condensation\cite{Sen:1998sm}.

It is easy to check that the above may not be readily realizable in supergravity. For example, if one looks at the dilaton equation of motion in type IIB, because the dilaton field couples non-trivially to the DBI action of the D$9$ (or $\overline{{\rm D}9}$) brane, there is no constant dilaton solution with only a five-form flux in the background. At best, one may hope there is a regime of slowly varying dilaton field, in which the action that we study, may closely approximate such a configuration in supergravity. We will not elaborate more on this issue in the current article and leave this for future explorations.

Our goal here is more modest. We intend to explore the solution space of the proposed action. The primary motivation for the inclusion of a Chern-Simons term can be drawn from the observation made in \cite{Kundu:2017cfj}, in which it was demonstrated that, by tuning the parameter $\left(N_f/N_c \right)$, one may interpolate between a spatial modulation instability and a superconducting instability, similar to the physics of QCD at large $N_c$ and large $N_f$\cite{Shuster:1999tn}. Thus, we will treat the Newton's constant, the putative brane tension and the Chern-Simons coefficients which appear in defining the action of the theory to be free, and we will explore the nature of various exact analytical solutions, as one varies these parameters. In the putative dual, these correspond to the the rank of the gauge group, $N_c$, the number of the fundamental matters, $N_f$, and an anomaly-induced transport coefficient.

We will work in $(4+1)$-bulk dimensions, and subsequently seek analytical solutions of the gravity equations of motion with a certain flux excited on the DBI-sector. This particular flux ansatz corresponds to having a non-vanishing (constant) magnetic field and a non-vanishing density (therefore, a non-vanishing chemical potential) in the boundary dual theory. A priori, one can think of starting with an AdS$_5$ asymptotic, which we show is preserved by the various fields that we turn on, and reach an appropriate IR geometry. This should correspond to a renormalization group (RG) flow, from an UV $(3+1)$-dimensional conformal field theory (CFT) to an IR CFT. In the generic cases, though, we will find that an explicit interpolating geometry in gravity may be subtle to obtain, specially in the case of vanishing temperature. Moreover, in the presence of both density and magnetic field, the IR is described by an warped AdS$_3 \times R^2$ geometry, or a Schr\"{o}dinger symmetric solution in three bulk dimension, denoted by Schr$_3 \times R^2$. Generically, the role of the magnetic field, in the IR, is to decouple the $R^2$ (or a $T^2$),\footnote{Note that, this can already be seen in holography with explicit D-brane models, where the flavour degrees of freedom are treated as probes in the system; see {\it e.g.}~\cite{Filev:2007gb}-\cite{Alam:2012fw}. Similar probe analysis with non-vanishing density is also analyzed widely, see {\it e.g.}~\cite{Kobayashi:2006sb, Mateos:2007vc}.} from the rest of the dynamics; see {\it e.g.}~\cite{Jain:2014vka} for a similar phenomenon. Both the warped AdS$_3$ (henceforth denoted by wAdS$_3$) and the Schr$_3$ geometries are non-perturbative in $\left( N_f / N_c\right) $, as well as the Chern-Simons coefficient, which we denote by $\alpha$.\footnote{The large fundamental matter number, or the so-called Veneziano limit has been explored in details in the literature, see {\it e.g.}~\cite{Alho:2013hsa}-\cite{Gursoy:2016ofp}. Our focus is complimentary to those.}

The magnetic field breaks Lorentz invariance SO$(1,3) \to {\rm SO}(1,1) \times {\rm SO}(2)$, which is further broken by the density down to a scaling symmetry. The Schr\"{o}dinger symmetry can be accompanied by a special conformal transformation for a certain case, but generically no such symmetry is preserved. From this perspective, it is natural to expect the emergence of a wAdS$_3$ or a Schr$_3$ geometry in the IR. The parameters of our action, namely $N_f / N_c$ and $\alpha$, and the parameters of the solution, namely the magnetic field and the density, ultimately yields an infinite class of both these solutions. Warped AdS solutions are present only within a subset of the parameter space of the action, in which Schr\"{o}dinger solutions always exist. Thus, within the overlap region, there may be a competition between the two candidate IR descriptions.

Since the Chern-Simons term qualitatively emulates a triangle anomaly in the dual field theory, and we work in a limit where $ \left( N_f / N_c \right)$ is not necessarily small, this effect is certainly non-trivial in the IR.\footnote{Note that, typically, U$(1)$ anomalies are suppressed at large $N$.} In a certain sense, anomalies are understood as the IR physics of a system, see {\it e.g.}~\cite{Harvey:2005it}. Information of anomalies are encoded in the analytic structures of current-current correlation functions. Specifically, discontinuities of such correlation functions at vanishing momenta are directly related to non-conservation of a classical current. The importance of vanishing momenta clearly indicates the IR dynamics of the system. For a QCD-type theory, including the effect of such anomalies in the low energy dynamics is also of phenomenological interest, specially in the presence of a density and a magnetic field, see {\it e.g.}~\cite{Fukushima:2008xe}, or see {\it e.g.}~\cite{Hoyos:2011us} for a holographic perspective. Within our model, such effects can be explored in details, and beyond the so-called probe approximation $N_f \ll N_c$ limit. We leave this for future explorations.

\begin{figure}[ht!]
\begin{center}
{\includegraphics[width=0.8\textwidth]{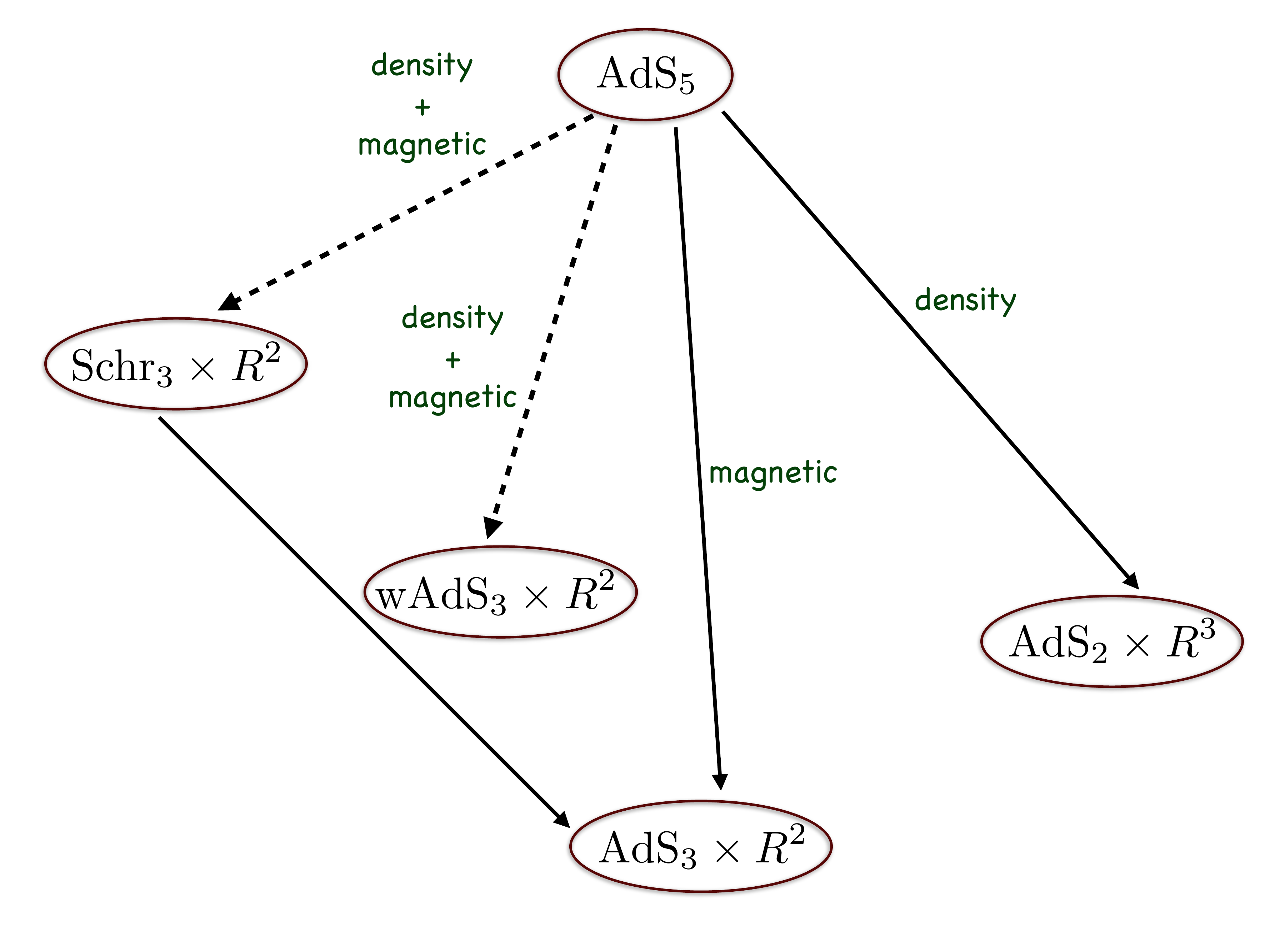}}
\caption{\small A schematic form of the solution space. The solid arrows represent the cases where explicit flows have been constructed, see in \cite{Kundu:2016oxg}. The dashed arrows are, at least, not explicitly known and represents a plausible scenario. Currently, we have no {\it a priori} mechanism of distinguishing between the flow to the Schr\"{o}dinger geometry and the warped AdS geometry.} \label{interpolate}
\end{center}
\end{figure}

Early studies on QCD-like theories with a vanishing beta function, as one tuned the number of adjoint and fundamental degrees of freedom, can be summarized by the so-called Caswell-Banks-Zaks fixed point\cite{Caswell:1974gg, Banks:1981nn}. Our analysis, in this article, can be viewed along a similar direction, where the effect of additional fields, such as density or a magnetic field, has been taken into account in the presence of triangle anomalies. The IR {\it fixed points} have been analyzed on their own rights, specially the wAdS$_3$ geometry. We have summarized the results, pictorially, in fig.~\ref{interpolate}.
The pictorial representation is based on our analysis of linearized equations around each {\it fixed point}. These are summarized in five appendices (\ref{appen1})-(\ref{appen5}). However, the linearized analyses cannot distinguish between the flow to the Schr\"{o}dinger and the warped AdS geometry.

It turns out that one can introduce an event horizon for the wAdS geometries, and subsequently carry out a {\it local} version of holography for such states, without worrying about the UV-completion of the system. At sufficiently low temperature, we demonstrate this in the main text, one obtains a negative charge susceptibility for the wAdS$_3$ black hole geometry.\footnote{The specific heat is always negative for these black holes, which also indicates a thermodynamic instability.} This signals an instability towards the formation of a crystalline phase for the system, similar to the results of \cite{Faedo:2017aoe} where an explicit D-brane construction has been considered. We have not managed to find a black hole solution for the Schr$_3$ geometry, thus we are unable to comment on that.

The emergence of an wAdS$_3$ geometry, and correspondingly an warped dual CFT, in the context of a phase for QCD with a non-vanishing density and a background magnetic field, is an interesting aspect in itself. One may explore how the SL$(2, R) \times $U$(1)$ isometry of the background gets extended to a complete Virasoro-Kac-Moody, within the action that we study. This is likely to constrain boundary conditions for the matter fields that we have introduced and it will be an interesting exercise to explore that.\footnote{We thank Nemani Suryanarayana for conversation on this point.} Moreover, the possibility of a wCFT playing a pivotal role in QCD-like phenomenology, is enough reason to analyze wCFT in general, perhaps along the lines of recent studies in \cite{Song:2017czq}. We leave these and many other interesting aspects for future work.

This article is divided into the following sections: In section $2$, we begin with the model and the ansatz that we will explore. In the next section we discuss the data at the horizon, including the explicit solutions. In section $4$, we make a brief detour to connect our wAdS$_3$ solutions with the existing literature. Section $5$ is devoted to the discussion of thermodynamics of each individual IR descriptions, specially the ones for wAdS$_3$. Finally, details of the linearized analyses are provided in appendices (\ref{appen1})-(\ref{appen5}).

\section{The action \& the ansatz}

We will work with the following action in $(4+1)$- bulk dimensional space-time
\be \label{action1}
\begin{split}
S & =  S_{\rm grav} + S_{\rm DBI} + S_{\rm CS} \ , \\
S_{\rm grav} & =  {1 \over 2 \kappa^2} \int d^{5} x \sqrt{-g} ~(R-2 \Lambda) \ , \quad  S_{\rm DBI} = - \tau  \int d^{5} x \sqrt{-\text{det}~(g+F)} \ , \\
S_{\rm CS} & =  {\alpha \tau \over 3!} \int d^{5} x~ A \wedge F \wedge F \ .
\end{split}
\ee
Here $\kappa$ and $\tau$ are dimensionful couplings in the action, which yield one free dimensionless parameter $(\kappa^2\tau)$. Qualitatively, from an AdS/CFT perspective, one can identify $(\kappa^2\tau) \sim N_f / N_c$, where $N_f$ and $N_c$ are the numbers of fundamental and adjoint matter degrees of freedom, respectively. There is a negative cosmological constant, denoted by $\Lambda = - 6/ L^2$, where $L$ is an overall curvature scale. The cosmological constant typically arises from integrating supergravity fluxes on some compact manifold. The Dirac-Born-Infeld matter, which arises from the open string degrees of freedom, has a non-vanishing U$(1)$ gauge field, denoted by $F = dA$. We have also included a non-vanishing Chern-Simons term, with an additional dimensionless free parameter $\alpha$.

The equations of motion following from the action are as given below
\be \label{eoms1}
\begin{split}
& R_{\mu \nu} - {g_{\mu \nu}  \over 2} (R -2  \Lambda)  =  T_{\mu \nu} \ ,\\
& \partial_{\mu} \big(\sqrt{-\text{det}~(g+F)} \mathcal{A}^{\mu \nu} \big) + {\alpha \over 2} \epsilon^{\nu\alpha\beta\rho\sigma}F_{\alpha\beta}F_{\rho\sigma} = 0 \ , 
\end{split}
\ee
where 
\be \label{eoms2}
\begin{split}
\mathcal{A}^{\mu \nu} &= -\left( {1 \over g+F}~.~F~.~{1 \over g-F}\right)^{\mu\nu} \ , \\
T^{\mu \nu} &= {\kappa^2 \tau \over \sqrt{-g}} \left[{\delta S \over \delta g_{\mu \nu}} + {\delta S \over \delta g_{\nu \mu}} \right] = - (\kappa^2 \tau) {\sqrt{-\text{det}~(g+F)} \over \sqrt{-g}}S^{\mu \nu} \ , \\
\text{and} ~~ S^{\mu \nu} &= \left( {1 \over g+F}~.~g~.~{1 \over g-F}\right)^{\mu\nu} \ .
\end{split}
\ee

Following closely the treatment as well as the notation of \cite{D'Hoker:2009bc}, we will work with the following ansatz, in order to look for solutions of (\ref{eoms1})-(\ref{eoms2}):
 \be \label{ansatz1}
 \begin{split}
 F=& E(r) dr \wedge dt+ B \, dy^1  \wedge dy^2 + P(r) dx \wedge dr \ , \quad \vec{y} = \left\{ y^1, y^2 \right\} \ , \\
 ds^2 =& -U(r)dt^2 + {dr^2 \over U(r)} + e^{2V(r)} d\vec{y}^2 + e^{ 2W(r)} \left(dx + C(r)dt \right)^2 \ . 
\end{split}
 \ee

The general goal is to substitute the above ansatz for the field strength and the metric to the equations \eqref{eoms1}-(\ref{eoms2}) and solve the corresponding differential equations to obtain the geometric data: $\left\{E,P,U,V,W,C \right\}$, in terms of the radial co-ordinate, $r$.

\section{Horizon data and extremal solutions} 

Certainly, the general solution of (\ref{eoms1})-(\ref{eoms2}) is hard to obtain, thus we will focus on a certain simple yet interesting class of solutions. Our primary interest is to describe the infrared, equivalently, the low temperature physics. In the extreme limit, this corresponds to precisely the zero temperature physics, and therefore extremal black hole solutions in the bulk.

Towards that, let us suppose that the event horizon is located at $r=r_{\rm H}$. Before imposing the extremal condition, at the horizon, one imposes the following conditions:
\be 
 E\left(r_{\rm H} \right) = q \ , \quad  U \left(r_{\rm H} \right) = V \left(r_{\rm H}\right) = W \left(r_{\rm H}\right) = C\left(r_{\rm H}\right) = P\left(r_{\rm H}\right) = 0 \ .
 \ee
Note that, we have imposed more conditions that necessary. From the definition of event horizon, one obtains $U\left( r_{\rm H}\right) = 0 $, $V\left( r_{\rm H}\right) =0 = W\left( r_{\rm H}\right) $ amounts to an overall scaling of the $\vec{y}$ plane or the $x$-direction. One can also set $C\left( r_{\rm H}\right) = 0$ using a shift symmetry of our ansatz\footnote{The transformation $x \to x - \alpha t$, $C \to C + \alpha$, $E \to E - \alpha P$ is a symmetry of (\ref{ansatz1}), see \cite{D'Hoker:2009bc} for more details.}, however $P\left( r_{\rm H}\right) $ is a free parameter.

Given the above, the rest of the data at the horizon, $\left\{ U' \left( r_{\rm H}\right) , V'(r_{\rm H}),W'(r_{\rm H}),C'(r_{\rm H})\right\}$, are constrained {\it via} non-linear algebraic relations. Solving the Maxwell and Einstein's equations, one obtains:
 \be
\begin{split}
& q \left(- \frac{\sqrt{B^2+1}  }{\sqrt{1-q^2}}C'(r_{\rm H}) - 4 \alpha  B\right) = 0 \ , \\
& U'(r_{\rm H})W'(r_{\rm H}) =  -\frac{4 \left(\kappa ^2 \tau  \sqrt{\frac{1-q^2}{B^2+1}} \left(B^2 q^2+1\right)+\Lambda  \left(1-q^2\right)\right) + 3 \left(1-q^2\right)C'(r_{\rm H})^2}{6 \left(1-q^2\right)}  \ , \\
& U'(r_{\rm H})V'(r_{\rm H}) = - \frac{\kappa ^2 \tau  \sqrt{\frac{1-q^2}{B^2+1}} \left(2-B^2 \left(q^2-3\right)\right)-2 \Lambda  \left(q^2-1\right)}{3 \left(1 - q^2 \right)} \ . 
\end{split}
\ee
In what follows, we consider the specific value of $\Lambda= - 6$. Now, imposing the extremality condition, $U'(r_{\rm H})=0$, we can solve the resulting algebraic equations to obtain:
\be 
\begin{split}
& q = 0 \ , \quad B^2 = \frac{4 \left( \pm \sqrt{3 \kappa ^4 \tau ^2+36}+6\right)-2 \kappa ^4 \tau ^2}{3 \kappa ^4 \tau ^2} \ ,  \quad C'(r_{\rm H})^2 = 16 \mp 4 \sqrt{\frac{\kappa ^4 \tau ^2}{3} + 4} \ .  \\
&   B = 0 \ , \quad C'(r_{\rm H}) =  0 \ , \quad  q^2 = 1-\frac{\kappa ^4 \tau ^2}{\Lambda ^2} \ ; \quad {\rm or} \\
& B^2 = \frac{16 \alpha^2  (8 \alpha ^2+\Lambda)}{\kappa ^4 \tau^2 } \ , \quad C'^2(r_{\rm H}) = -4 (8 \alpha ^2+\Lambda ) \ , \quad q^2 = -\frac{\kappa ^4 \tau ^2}{64 \alpha ^4}+\frac{\Lambda }{4 \alpha ^2}+3 \ .
\end{split}
\ee
Any non-trivial solution must exist along one of the above algebraic solutions. The first two cases are not interesting since the Chern-Simons sector decouples. Let us discuss the second case.

Demanding consistency of the solutions ({\it i.e.}~imposing $B^2>0$, $q^2>0$ and $C'^2(r_{\rm H})>0$), we obtain the following allowed range for the parameters, corresponding to the conditions discussed above:  
\be 
\begin{split}
&  0<\kappa^2 \tau <4 \alpha  \sqrt{12 \alpha ^2-6}, \quad \frac{1}{\sqrt{2}}<\alpha <\frac{\sqrt{3}}{2} \quad {\rm or} \\
&  -4 \alpha  \sqrt{12 \alpha ^2-6}<\kappa^2 \tau  <0, \quad \frac{1}{\sqrt{2}}<\alpha <\frac{\sqrt{3}}{2} \ . 
\end{split}
\ee 
We will now present the explicit solutions that we analytically obtain.

We can consider the purely electric and the purely magnetic cases, in which the Chern-Simons coefficient plays no role. These are the solutions that were already obtained and analyzed in \cite{Kundu:2016oxg}.

\subsection{Vanishing magnetic field}

This extremal solution is simply an ${\rm AdS}_2 \times R^3$ geometry, described by the following functions:
\be 
U(r)=L_1 r^2 \ , \quad  E(r)=q \ ,\quad C(r)=W(r)=V(r)=P(r)=B=0 \ . 
\ee
The metric and the gauge field are given by
 \be 
 \begin{split}
F=& q~ dr \wedge dt \ , \quad ds^2 = - L_1 r^2 dt^2 + {dr^2 \over L_1 r^2}  +  d\vec{y}^2 + dx^2 \ ,
\end{split}
\ee
with
\be 
\Lambda = -\frac{L_1}{q^2} \ , \quad \kappa^2 \tau = \frac{\sqrt{1-q^2}}{q^2}L_1 \ . \label{ads2param}
\ee
%

\subsection{Vanishing electric field}

In this case, the extremal geometry is given by an ${\rm AdS}_3 \times R^2$ solution. In terms of our ansatz data, this is described by
\be 
U(r)=L_1 r^2 \ ,~~ W(r)= \log(r) \ ,~~C(r)=P(r)=E(r)=V(r)= 0 \ . \label{ads3sol1}
\ee
such that the metric and the gauge field take the form:
\be 
\begin{split}
F=& B~ dx_1  \wedge dx_2  \ , \quad ds^2 = - L_1 r^2 dt^2 + {dr^2 \over L_1 r^2}  +  d\vec{y}^2 + r^2 dx^2 \ . \label{ads3sol2}
\end{split}
\ee
The parameters of the solution are related to the parameters of the action by the following relations:
\be 
\Lambda =  -3L_1 - \frac{2L_1}{B^2} \ ,~~\kappa^2 \tau = \frac{2L_1\sqrt{B^2+1}}{B^2} \ . \label{ads3sol3}
\ee
%

\subsection{Non-vanishing electric \& magnetic field}

Recall that, when $\alpha =0$, {\it i.e.}~in the limit of vanishing Chern-Simons coupling, there is no scaling-type analytical solution with both electric and magnetic components excited. This was already observed in \cite{Kundu:2016oxg}. Introducing a non-vanishing Chern-Simons term, similar to the observation made in \cite{D'Hoker:2009bc}, allows one to obtain a scaling-type analytical solution. In fact, there are more than one such solution, which we now discuss. 

One family of solutions is given by an warped AdS$_3 \times R^2$ geometry. These are given by
\be 
U(r)=L_1 r^2 \ ,~~ C(r)= L_2~ r \ ,~~E(r)=q \ ,~~W(r)=V(r)=P(r)=0 \ . 
\ee
such that the two-form matter field and the metric data are give by
\be 
\begin{split}
F = & q dr \wedge dt + B \, dy^1  \wedge dy^2  \\
 ds^2 = & -L_1 r^2 dt^2 + {dr^2 \over L_1 r^2}  +  d\vec{y}^2 + (dx + L_2 r dt)^2 \ . \label{wads3}
\end{split}
\ee
The parameters of the solution are determined by the parameters of the theory, as follows:
\be \label{wads3para}
\begin{split}
L_2^2 = & -4 (8 \alpha^2+\Lambda ) \ ,~~B^2 = -\frac{16 \alpha^2 (8 \alpha^2+\Lambda )}{\kappa^4 \tau^2} \ , \\
L_1 = & -\frac{\kappa^4 \tau^2}{8 \alpha^2} - 8 \alpha^2 - 2 \Lambda \ ,~~q^2 = 3-\frac{\kappa^4 \tau^2-16 \alpha^2 \Lambda }{64 \alpha^4} \ .
\end{split}
\ee
As we have discussed before, consistency conditions impose further constraints on this parameter space. Here the constraints are obtained to be:
\be \label{wads3allowed}
\frac{\kappa^4 \tau^2}{16 \alpha^2}-12 \alpha^2<\Lambda \ , ~~\Lambda <-8 \alpha^2 \ ,~~\epsilon = \kappa^2 \tau \neq 0 \ .
\ee
A few comments are in order: First, as we had also observed in \cite{Kundu:2016oxg}, this new class of warped AdS$_3$ solutions are also non-perturbative in the fundamental matter back-reaction. Secondly, it is clear that, given $\epsilon = \kappa^2\tau$ and $\Lambda$, there is an infinite family of warped AdS$_3$ solutions, parametrized by the value of the Chern-Simons coupling $\alpha$. This is in stark contrast to \cite{D'Hoker:2009bc}, where only one such solution in the $\alpha$-space is allowed. The reason for this is intuitively simple: One has an additional natural parameter in the system, namely $\epsilon$, which is essentially the ratio of number of fundamental and adjoint matter. In fact, the solution space of the wAdS$_3$ geometry can be represented by a range of values for $\epsilon \left( \alpha, \ldots\right)$, for parameters defining the action.

There is another class of extremal solutions, namely the Schr\"{o}dinger geometries. To describe this class, we use the following form of the geometric ansatz:
\begin{eqnarray} \label{schrsoln}
ds^2 & =& -M(r) \left(dx^+\right)^2 + \frac{1}{L(r)^2}~ dr^2 + e^{2V(r)} d\vec{y}^2 + 2L(r)~ dx^+~dx^- \ , \\
F &=& E(r) dr \wedge dx^+ + B \, dy^1 \wedge dy^2 \ , 
\end{eqnarray}
The above solution is written in terms of null co-ordinates of AdS$_3$, $x^+$ and $x^-$, that, if we are to relate these null directions to a time-like and a space-like co-ordinate, can be given by: $x^+ = t + x$ and $x^- = (1/2) (x - t)$.\footnote{Note that, if we turn off the function $M$ and $E$, we indeed get the following geometry:
\begin{eqnarray}
ds^2 = L(r) \left( -dt^2 + dx^2 \right) +   \frac{dr^2}{L(r)^2} + d\vec{Y}^2  \ , \quad F = B \, dy^1 \wedge dy^2 \ ,
\end{eqnarray}
with $L(r) \sim r$ and thus the geometry is purely AdS$_3 \times R^2$.} Here, the various functions are obtained to be: 
\begin{eqnarray} \label{schrfunc}
L(r) &=& L_1 r \ , \quad V(r) = \frac{1}{2} \log \left(L_2 \right) \ , \quad E(r) = q \, r^{ - \xi - 1} \ , \\
M(r) & = & M_0 \, r + \frac{ L_1^2 L_2 q^2  \left(B^2 + L_2^2\right){}^{3/2} }{4 \alpha  B \left(L_1 L_2 \sqrt{B^2+L_2^2}+8 \alpha  B\right)} \left( \kappa ^2 \tau \right)  r^{- 2\xi } \ .
\end{eqnarray}
The constant $M_0$ is undetermined. The map between the parameters of the action to the parameters of the solution is explicitly given by
\begin{eqnarray} \label{schrsolnpara}
&& \Lambda = -\frac{L_1^2 \left(3 B^2+2 L_2^2\right)}{4 B^2} \ , \quad \kappa^2 \tau = \frac{L_1^2 L_2 \sqrt{B^2+L_2^2}}{2 B^2} \ , \\
&& \xi = \frac{4 \alpha B}{L_1 L_2 \sqrt{B^2 + L_2^2}} \ .
\end{eqnarray}
Using rescaling symmetry of the $\vec{y}$-plane, we can set $L_2 =1$. This still leaves us with two independent parameters, in the solution, namely $L_1$ and $B$, as compared to \cite{D'Hoker:2009bc} which has only one independent parameter. Thus, we have an infinite family of the Schr\"{o}dinger geometries, parametrized by {\it e.g.}~the magnetic field. Equivalently, one obtains an infinite family of boundary field theories characterized by a free dynamical exponent. Furthermore, note that the exponent $\xi$ is a function of both the Chern-Simons coupling, as well as the magnetic field. Thus, with a fixed set of parameters defining the theory ({\it i.e.}~with a fixed value of the Chern-Simons coefficient), one obtains an infinite family of solutions, in which the functional behaviour of $M(r)$ differs significantly.

Let us offer some comments regarding the geometry in (\ref{schrsoln}). The simplest way to see the Schr\"{o}dinger symmetry is to choose the radial gauge $\rho^2 = r$ and check explicitly the invariance of the data in (\ref{schrsoln}), under the transformation:
\begin{eqnarray}
x^+ \to \lambda^{1+\nu} x^+ \ , \quad x^- \to \lambda^{1-\nu} x^- \ , \quad \rho \to \lambda^{-1} \rho \ ,
\end{eqnarray}
where $\lambda$ is a real constant, $\nu$ is a real number and the corresponding dynamical exponent is given by $(\nu+1)$. It is easy to check that $\nu = - 2 \xi - 1$. Interestingly, the specific solution in (\ref{schrfunc}) requires $M(r) > 0$, which, in the limit $M_0 = 0$ implies:
\begin{eqnarray}
\alpha B \left(L_1 L_2 \sqrt{B^2+L_2^2}+8 \alpha  B\right) > 0 \ , 
\end{eqnarray}
The above condition is trivially satisfied in the range $\alpha B >0$, in which the allowed values are $\xi \in \left[ 0, \infty\right]$ and therefore $\nu \in \left[ -1, -\infty\right]$. However, in the regime $\alpha B < 0$, the constraint is non-trivial and further requires: 
\begin{eqnarray}
\left | 8 \alpha B \right | > L_1 L_2 \sqrt{B^2+L_2^2} \quad \implies \quad \left | 2 \xi  \right | > 1 \quad \implies \quad \nu > 0 \ .  
\end{eqnarray}
The $\left|2\xi \right| = 1 $ point is certainly included in within the $\alpha B >0$ branch. Also, the special case of $\nu = 1$, which enhances the corresponding symmetry group by including a special conformal symmetry, is located at the point $\xi = 1$, in the $\alpha B  < 0$ regime of the parameter space. Thus the allowed values of the dynamical exponent are given by $
\nu \in \left[ -1 , - \infty \right] \cup \left[0, \infty \right]$.

Before discussing further the infrared solutions, specially the wAdS$_3$ background, we will make a short digression on warped AdS$_3$-geometries, to connect with the discussion in \cite{Anninos:2008fx}.

\section{Warped AdS solutions}

The pure AdS$_3$-geometry can be seen as a Hopf fibration on a Lorentzian or an Euclidean AdS$_2$, by a spacelike or a timelike fibration. Thus, correspondingly, the pure AdS$_3$ geometry can be written as, using the notation of \cite{Anninos:2008fx}:
\begin{eqnarray}
ds^2 & = &  \frac{\ell^2}{4} \left[ - \cosh^2 \sigma d\tau^2 + d\sigma^2 + \left( du + \sinh\sigma d\tau\right) ^2\right] \\
& = & \frac{\ell^2}{4} \left[ \cosh^2\sigma du^2 + d\sigma^2 - \left( d\tau + \sinh\sigma du \right)^2\right] \ . 
\end{eqnarray}
The warped AdS$_3$ is simply obtained by fibration with a non-trivial factor. This non-trivial factor reduces the ${\rm SL}(2,R) \times {\rm SL}(2,R)$ symmetry of the AdS$_3$ down to ${\rm SL}(2,R) \times {\rm U}(1)$. Depending on the nature of the fibration, the warped AdS$_3$ can be divided into three sub-classes.

\subsection{Spacelike warped solution}

We adopt a similar notation for our case. The spacelike warped ${\rm AdS}_3 \times R^2$ solution is given by 
\begin{equation}
ds^2 = \ell \left[- \cosh^2 \sigma ~dt^2  + d\sigma^2 + L(dx + \sinh\sigma~ dt)\right]^2 + d\vec{Y}^2 \ ,
\end{equation}
where $L$ is the warping factor. We will set $\ell=1$ for rest of discussion. Upon setting $\sinh\sigma = r$  we get:
\begin{equation}
ds^2 =  \left[- (1+ r^2)  dt^2  + \frac{dr^2}{1+r^2} +  L(dx + r dt)^2\right] + d\vec{Y}^2 \ . 
\end{equation}
The above is a solution to our equations of motion along with gauge field ansatz $E(r) =q,~ B= {\rm const.}$, with
\begin{eqnarray}
 \nonumber L &=& \frac{2 B^2 \left(1-q^2\right)}{B^2 \left(2-q^2\right)+q^2} \ ,~~ \Lambda = -\frac{B^2 \left(3-q^2\right)+2}{2 \left(B^2 \left(2-q^2\right)+q^2\right)} \ , \\
 \kappa^2 \tau & = & \frac{\sqrt{\left(B^2+1\right) \left(1-q^2\right)}}{B^2 \left(2-q^2\right)+q^2}\ ,  \quad \alpha = -\frac{B \sqrt{\left(B^2+1\right) \left(1-q^2\right)}}{2 \left(B^2 \left(2-q^2\right)+q^2\right)} \ .
\end{eqnarray}
Clearly for $q^2<1$, $L<1$, so we only have spacelike squashed solution. Now this same set of these parameters also allow a Poincar\'{e} version of this solution that we shall mostly use in this paper, namely:
\begin{equation}
ds^2 =  \left[-  r^2  dt^2  + \frac{dr^2}{r^2} +  L(dx+ r dt)\right] + (d\vec{Y})^2 \ . \label{wadssquash}
\end{equation}
Note that (\ref{wadssquash}) is related to (\ref{wads3}) {\it via} $L \rightarrow L_2^2\, ,~ x \rightarrow L_2~ x$.

\subsection{Timelike warped geometries}

The timelike warped ${\rm AdS}_3 \times R^2$ solution is given by
\begin{equation}
ds^2 =  \left[ \cosh^2 \sigma~ dx^2  + d\sigma^2 - L(dt + \sinh\sigma~ dx)^2\right] + d\vec{Y}^2 \ .
\end{equation}
Equivalently, upon substituting $\sinh\sigma = r$ we get:
\begin{equation}
ds^2 =  \left[ (1+r^2) dx^2  + \frac{dr^2}{1+r^2} - L(dt + r dx)\right]^2 +  d\vec{Y}^2 \ .
\end{equation}
We do not have these solutions.

\subsection{Null warped solutions}

Finally, the null-warped ${\rm AdS}_3 \times R^2$ solution, in the notation of \cite{Anninos:2008fx}, is given by
 \begin{equation}
 ds^2 = \left[ \frac{du^2 + dx^+ ~ dx^-}{u^2} \pm {\frac{\left(dx^- \right)}{u^4}}^{2} \right] + d\vec{Y}^2 \ , \label{nullwads}
 \end{equation}
where $x^{\pm}$ are null coordinates constructed out of $\{t,~ x\}$. Upon substituting $u= \frac{1}{\rho}$ followed by $\rho^2 = r$ , we get obtain the following form: 
\begin{equation}
ds^2 = \left[ \frac{dr^2}{4r^2} + r dx^+~dx^-  \pm r^2 (dx^-)^2\right] + d\vec{Y}^2 \ .
\end{equation}
This is a solution with:
\begin{eqnarray}
\nonumber \Lambda &=& -\frac{2}{B^2}-3 \ , \hspace{4mm} \kappa^2 \tau = \frac{2 \sqrt{B^2+1}}{B^2} \ ,
\end{eqnarray}
which corresponds to the Schr$_3$-geometry discussed in (\ref{schrsoln}) with dynamical exponent two. For a general dynamical exponent the corresponding null deformation in (\ref{nullwads}) will take a form $\left(dx^- \right)^2 / u^{2n}$, where $n$ is the corresponding dynamical exponent.


\section{Physics at the infra-red} 

In this section, we intend to view the various (extremal) solutions having a holographic description in their own terms, irrespective of whether one is able to geometrically construct smooth interpolation to an asymptotically AdS$_5$ geometry. In this sense, we remain oblivious to any potential UV-completion of the infra-red physics, which is anyway the limitation of our approach, since it lacks a precise description in terms of branes and strings. In particular, we will probe one particular potential instability, which is captured by computing the charge susceptibility of a thermodynamic description of the corresponding dual field theory. Towards that, one needs to first incorporate a {\it small} temperature and perturb away from extremality. This temperature, for us, is a placeholder for reliably carrying out thermodynamics by computing the on-shell renormalized Euclidean action for each solution.

\subsection{Non-extremal Warped ${\rm AdS}_3 \times R^2$}

This particular solution is rather unique and, in this section, we will discuss some thermodynamic properties of the corresponding state. The non-extremal ${\rm wAdS}_3 \times R^2$ solution is given by
\begin{eqnarray}
ds^2 &=& - L_1 r^2 ~ f(r) dt^2 + \frac{dr^2}{L_1 r^2 ~ f(r)} +  d\vec{Y}^2 + \left( dx_3 + L_2 r dt \right)^2 \ , \label{wadsbh1} \\
F &=& q~ dr \wedge dt + B~  dy^1 \wedge dy^2 \ , \label{wadsbh2}
\end{eqnarray}
where
\begin{eqnarray}
f(r)& =& 1+\frac{c_1}{r}+\frac{c_2}{r^2}, \quad \Lambda = -\frac{4 \alpha ^2 \left(B^2 \left(3-q^2\right)+2\right)}{B^2+1} \ , \quad \tau =\frac{8 \alpha ^2 \sqrt{\frac{1-q^2}{B^2+1}}}{\kappa ^2} \ , \label{wadsbh3} \\
L_2 &=& -4 B \sqrt{\frac{1-q^2}{B^2+1}} \alpha , \quad  L_1 = -\frac{8 \alpha ^2 \left(B^2 q^2-2 B^2-q^2\right)}{B^2+1} \ . \label{wadsbh4}
\end{eqnarray}
Without losing any generality, we can set the horizons at $r=0$ and $r=1$, respectively. This yields: $f(r)= 1-\frac{1}{r}$, with an associated temperature $T = \frac{L_1}{4\pi}$, which is determined by the outer horizon. Ignoring now the possibility of constructing an explicit RG-flow to connect this solution to an asymptotic AdS$_5$-background, we will explore the thermodynamic properties of this particular putative IR state. For this purpose, we assume that it makes sense to define holography in such a scenario, since the IR can exist in its own right without explicit reference to a possible UV-completion. Towards that, we will carry out the usual holographic prescription in deriving the thermodynamic free energy, by calculating on-shell Euclidean action; and in extracting the stress-tensor of the dual field theory, by performing a Lorentzian analysis of the Brown-York quasi-local tensor.

To describe the Euclidean on-shell action, one requires to have a real Euclidean section of our Lorentzian solution given in (\ref{wadsbh1})-(\ref{wadsbh4}). This is obtained by performing the following analytic continuation:
\begin{eqnarray}
t \to - i t_{\rm E} \ , \quad q \to i Q \ , \quad L_2 \to  i L_2^{\rm E} \ .
\end{eqnarray}
The corresponding Euclidean on-shell action is given by
\begin{eqnarray}
\mathcal{S}_{\rm on-shell}^{\rm E} & = & - \frac{1}{2\kappa^2}\int_{M} d^5x\sqrt{g}(R-2\Lambda) -  \frac{1}{\kappa^2}\int_{\partial M}d^4 y \sqrt{h} K  + \tau \int_{M} d^5x\sqrt{\left| g+F \right|}   \nonumber\\  
&- & \xi_1 \int_{\partial M}d^4 y \sqrt{h} -  \xi_2 \int_{\partial M}d^4 y \sqrt{h} h^{ab}n^r A_aF_{rb} \ . \label{euonshell}
\end{eqnarray}
In the above, $M$ denotes the full bulk spacetime over which the integration is carried out. The second term, containing $\{h ,K\}$, which are yet to be defined by the following data:
\begin{eqnarray}
&& n_{\mu} = \left(\sqrt{g_{rr}},0,0,0,0 \right) \ , \\
&& K_{\mu\nu}= \nabla_{\mu}n_{\nu}  \ ,\quad K_{ab} = \varphi^* \left( K_{\mu\nu} \right)  \ , \quad K = h^{ab}K_{ab} \ ,
\end{eqnarray}
is the Gibbons-Hawking term, with $h_{ab}$ as the induced metric on an $r={\rm const}$ slice. The two terms in the second line of (\ref{euonshell}), which are both boundary terms, are introduced to remove the divergence of the on-shell action evaluated on the wAdS$_3$ solution.\footnote{Certainly, we imagine a scenario where, the UV of the theory can be characterized by a cut-off. However, all physical answers should remain independent of this cut-off.} Note that, we have introduced two counter-terms, characterized by $\xi_{1,2}$, since both of these contribute to the same order in divergence and we have no {\it a priori} reason to rule any of them out.\footnote{Of course, one can introduce many more counter-terms, progressively complex in their structure. However, we are explicitly excluding those.}

Now cancellation of all divergence of (\ref{euonshell}), on-shell, requires:
\begin{equation}
\xi_1 = \frac{-4 \left(B^2+1\right) \xi_2 \kappa ^2 Q^2+\left(B^2+1\right) \left(L_2^{\rm E}\right)^2+16 \alpha ^2 \left(B^2 \left(Q^2-1\right)+2 Q^2\right)}{4 \left(B^2+1\right) \kappa ^2 \sqrt{L_1}} \ .
\end{equation}
which yields:
\begin{equation}
\mathcal{S}_{\rm on-shell}^{\rm E} = \left[ -\frac{\xi_2 Q^2}{2}-\frac{4 \alpha ^2 B^2 \left(Q^2+1\right)}{\left(B^2+1\right) \kappa ^2}  \right] \frac{1}{T} \ .
\end{equation}
As we can explicitly see, the constant $\xi_2$ remains undetermined.

To fix this, we can derive the boundary field theory stress-tensor, by evaluating the Brown-York tensor, in the Lorentzian picture. This is given by 
\begin{eqnarray}
T_{ab} & = & - \frac{2}{\sqrt{-h}} \frac{\delta S} {\delta h^{ab}} \nonumber\\
& = & - \frac{1}{\kappa^2} \left( K_{ab} - h_{ab}~K \right ) + \xi_1 h_{ab} + \xi_2 \left(h_{ab} h^{cd} n^{r} A_{c} F_{rd} - 2 n^{r} A_{a} F_{rb} \right) \ .
\end{eqnarray}
Given the formula above, it is straightforward to evaluate $T_b^a$ components, and finally, use impose the condition: $T^{x}_{x} = P =  - T \mathcal{S}_{\rm on-shell}^{\rm E}$, we get
\begin{equation}
\xi_2 = \frac{4 \alpha  B^2 \left(q^2-1\right) \left(-\frac{2 \alpha }{B^2+1}-\frac{\sqrt{2} q^2 \sqrt{\frac{B^2+1}{q^2-B^2 \left(q^2-2\right)}}}{B^2 \left(q^2-2\right)-q^2}\right)}{\kappa ^2 q^2} \ . 
\end{equation}
In the above, we have assumed that the on-shell Euclidean action yields the Gibbs free energy of the system, which, in turn, is given by the negative of the pressure.

Finally we can calculate charge susceptibility, we use the following:
\begin{eqnarray}
 && T^{t}_{t} = - E \ , \quad G = - P \ , \quad s = - \frac{\partial G}{\partial T} = - T \frac{\partial \mathcal{S}_{\rm on-shell}^{\rm E}}{\partial T} - \mathcal{S}_{\rm on-shell}^{\rm E} \ , \\
 &&  \mu = \frac{E - T s - G}{\rho} \ , \quad \rho = \frac{\delta S}{\delta F_{rt}} \ . 
\end{eqnarray}
Here $E$, $G$, $s$ are the energy, Gibbs free energy and entropy of the putative dual field theory, respectively; $T$ is the temperature and $ \rho $, $\mu$ are the boundary density operator and the chemical potential and $S$ is the Lorentzian on-shell action, respectively.

Now all the above quantities will be functions of the solution parameters $q$ and $B$. To realize them as purely field theoretic quantities we can express them as functions of $T$ and $\rho$ inverting the relations:
\begin{equation}
T = \frac{L_1}{4\pi} \ , \quad \rho = -\frac{4 q \alpha^2}{\kappa^2} \ .
\end{equation}
This yields:
\begin{equation}
B = \sqrt{\frac{\kappa ^4 \rho ^2-8 \pi  \alpha ^2 T}{-32 \alpha ^4+\kappa ^4 \rho ^2+8 \pi  \alpha ^2 T}}, ~~~~~ q = -\frac{\kappa ^2 \rho }{4 \alpha ^2} \ .
\end{equation}

Finally, to understand stability properties of the IR, we note that entropy and the specific heat are given by
\begin{eqnarray}
&& s = \frac{3 \kappa ^6 \rho ^4-8 \pi  \alpha ^2 \kappa ^2 \rho ^2 T}{128 \pi ^{3/2} \alpha ^4 T^{5/2}} \ , \\
&& C_{\rho}= T \left(\frac{\partial s}{\partial T}\right)_{\rho} = \frac{3 \kappa^2 \rho^4}{32\sqrt{\pi} \alpha^2 T^{\frac{5}{2}}}\left(\frac{T}{\rho^2}-\frac{5\kappa^4}{8\pi \alpha^4}\right) \ .
\end{eqnarray}
In the regime where entropy is strictly positive, one therefore finds: $C_\rho < 0$. This signals an instability. Furthermore, we calculate the charge susceptibility, given by
\begin{equation}
\chi^{-1} =\left( \frac{\partial \mu}{\partial \rho}\right)_T \ .
\end{equation} 
The exact expression is analytic, but not illuminating. In appropriate units, the small $T$ behaviour is given by
\begin{equation}
\chi^{-1}_{T \rightarrow 0} = -\frac{3 \left(\kappa ^6 \rho ^2\right)}{32 \left(\pi ^2 \alpha ^4\right) T^2}+\frac{\kappa ^2}{4 \pi  \alpha ^2 T}+\frac{64 \pi  \alpha
   ^2 \kappa ^2-3 \kappa ^6 \rho ^2}{128 \pi ^{3/2} \alpha ^4 \sqrt{T}}+\frac{\sqrt{T} \left(\frac{32 \pi }{\rho ^2}-\frac{\kappa
   ^4}{\alpha ^2}\right)}{16 \sqrt{\pi } \kappa ^2}+ \mathcal{O}\left(T^{5/2}\right) \ ,
\end{equation}
whereas the large $T$ behaviour is
\begin{equation}
\chi^{-1}_{T \rightarrow \infty} =\frac{2 \sqrt{\pi } \sqrt{T}}{\kappa ^2 \rho ^2}+\frac{9 \kappa ^2 }{16 \sqrt{\pi } \alpha ^2 \sqrt{T}} + \frac{\kappa ^2}{4 \pi \alpha ^2 T}-\frac{9 \left(\frac{1}{T}\right)^{3/2} \left(\kappa ^6 \rho ^2\right)}{128 \left(\pi ^{3/2} \alpha ^4\right)} - \frac{3 \left(\kappa ^6 \rho ^2\right)}{32 \left(\pi ^2 \alpha ^4\right) T^2} + \mathcal{O}\left(T^{-5/2}\right) \ .
\end{equation}
It is evident from above that, the presence of a non-vanishing $\rho$, in the small temperature limit, immediately guarantees an instability towards a clumping of charged objects. However, at large temperature, this instability goes away. The generic pattern of this instability in the full $\{\rho, T\}$-plane is, of course, richer. This is, qualitatively, similar to the quark-matter crystal phase hinted in \cite{Faedo:2017aoe} and perhaps also similar to \cite{Jensen:2017tnb}.

Let us conclude this discussion with a disclaimer. We have not managed to find an explicit flow to this IR, starting from an AdS$_5$ UV. We have discussed the technical complications towards this in appendix \ref{appen4}, which was also pointed out in \cite{D'Hoker:2009bc}. Essentially, the fluctuation modes, at vanishing temperature, always carry a relevant mode that grows unboundedly in the IR and therefore the wAdS becomes impossible to reach. Of course, one can avoid this problem by turning on a temperature; the other possibility is relaxing our ansatz. This is outside the scope of this work, and we will not explore this any further in the current work.

\subsection{Non-extremal ${\rm AdS}_2 \times R^3 $ }

The non-extremal ${\rm AdS}_2 \times R^3$ solution is given by
\begin{eqnarray}
ds^2 & = & - L_1 r^2 ~ f(r) dt^2 + \frac{dr^2}{L_1 r^2 ~ f(r)} + L_2 d\vec{X}^2 \ ,  \\
F &=& q~ dr \wedge dt \ , \quad f(r) =1 + \frac{c_1}{r} + \frac{c_2}{r^2} \ . 
\end{eqnarray}
where $c_{1,2}$ are arbitrary constants. In general, there are two roots of the function $f(r)$, corresponding to two horizons. Without any loss of generality, we can set the horizons to be at $r=0$ and $r=1$, which yields $f(r)= 1- \frac{1}{r}$. The corresponding Hawking temperature is given by $T= \frac{L_1}{4\pi}$. The parameters $\Lambda , \tau$ are already given in (\ref{ads2param}).

To discuss the regularisation, let us note that one add the following counter-terms:
\begin{eqnarray}
&& S_{\rm ct}^{(1)} =  c_1 \int_{\partial M}d^4 y ~\sqrt{h} \ , \\
&& S_{\rm ct}^{(2a)} = c_2 \int_{\partial M}d^4 y ~\sqrt{h} ~h^{ab}~n^r~ A_a F_{rb} \\
&& S_{\rm ct}^{(2b)} = c_3 \int_{\partial M}d^4 y ~\sqrt{h} ~h^{ab}~ A_a A_b \ .
\end{eqnarray}
The simplest choice is to use $S_{\rm ct}^{(1)}$, which eventually yields:
\begin{eqnarray}
\chi^{-1} = - \frac{2\sqrt{\pi ~ T}}{\kappa^2 \rho^2} < 0 \ ,
\end{eqnarray}
which signals a potential instability similar to the one observed in wAdS-background.

Before concluding this section, we note that the other counter-terms result in unphysical thermodynamic result, since one obtains a negative entropy which is derived from the corresponding free energy. The counter-term that we have used yields a vanishing entropy. One may object to this, since we know AdS$_2$ does indeed have a ground state degeneracy, which is also seen as the transverse $R^3$ on the event horizon. However, if we attempt to do holography which is decoupled from this transverse $R^3$, one naturally concludes a vanishing area of the corresponding event horizon. Essentially, this is equivalent to stating that, although one has a black hole solution, there is no non-trivial energy excitation on the AdS$_2$ throat, which is a familiar property. Note that, in \cite{Faedo:2014ana} a particular D-brane configuration was discussed, which yields to an AdS$_2$ throat supported by the two-form flux of the kind we have discussed here, for which one can compute the thermodynamics reliably. Such AdS$_2$ throats do not decouple from the UV-dynamics and therefore thermodynamic quantities such as entropy may be subtle to compute. Specifically, the approach pioneered in {\it e.g.}~\cite{Sen:2007qy} may prove instrumental towards that. We leave this for future explorations.

\subsection{Non-extremal ${\rm AdS}_3 \times R^2$}

It is simple to introduce an event-horizon for the IR described in (\ref{ads3sol1})-(\ref{ads3sol3}). The corresponding non-extremal ${\rm AdS}_3 \times R^2$ solution is given by rotating BTZ
\begin{eqnarray}
ds^2 &=& - \frac{(r^2 - r_+^2)(r^2 - r_-^2)}{L_1^2~r^2}  dt^2 + \frac{ L_1^2 ~r^2 dr^2}{ (r^2 - r_+^2)(r^2 - r_-^2) } +  d\vec{Y}^2 + r^2 \left( dx - \frac{r_+ r_-}{L r^2}dt\right)^2 \ , \nonumber\\
F &=& B~ dy^1 \wedge dy^2 \ .
\end{eqnarray}
Here, $r_{\pm}$ are the locations of the event horizons. Setting $r_-=0$, one obtains the standard BTZ black hole solution. The parameters of the action and the parameters of the solution are related by (\ref{ads3sol3}).

\section{Acknowledgements}

We thank Amit Ghosh and Nemani Suryanarayana for various conversations. We specially thank Nilay Kundu for numerous illuminating discussions and collaborations at the early stages of the work. We also thank the Department of Atomic Energy for funding. Finally, we humbly acknowledge the people of India for their support in research in basic sciences.

\appendix

\section{Appendix 1: Perturbing AdS$_5$ with various fields} \label{appen1}

In this appendix, we collect results from linearized analysis around the AdS$_5$-asymptotics, by various non-vanishing fields that we have in our ansatz in (\ref{ansatz1}). This exercise, among other things, readily captures the nature of the deformation to the putative conformal field theory at the boundary. For completeness, we also include the results that were already obtained in \cite{Kundu:2016oxg} for purely electric and purely magnetic perturbations.

In the absence of any DBI-flux, the solution is characterized by a negative cosmological constant: $\Lambda = - 6 / L_1  - \kappa^2 \tau $. One can consider linear fluctuations and solve Einstein equations to obtain:
\begin{eqnarray}
&& g_{tt} = r^2 L_1 \left(1 + \delta g_{tt} \right)  \ , \quad g_{xx} = r^2 L_1 \left(1 + \delta g_{xx} \right)  \ , \quad g_{rr} = r^{-2} L_1 \left(1 + \delta g_{rr} \right)  \ , \label{deltag1} \\
&&  \delta g_{tt} = \frac{\varepsilon}{r^4} \ , \quad \delta g_{xx} = \frac{p}{r^4} \ , \quad \delta g_{rr} = 0 \ , \quad {\rm with} \quad \varepsilon + 3 p = 0  \ . \label{deltag2}
\end{eqnarray}
The algebraic relation involving $\varepsilon$ (the energy) and $p$ (the pressure) is a simple consequence of tracelessness of the CFT energy-momentum tensor.

Now, a purely magnetic perturbation, sourced by turning on the flux $\delta F = Q_m dy^1 \wedge dy^2$ induces the following correction to the geometry:
\begin{eqnarray}
&& \Lambda = - \frac{6 + L_1 \kappa^2 \tau}{L_1} \ , \\
&& g_{tt} = L_1 r^2 \left(1 + \delta g_{tt} \right) \ , \quad g_{xx} = L_1 r^2 \left( 1 + \delta g_{xx}\right) \ , \quad g_{yy} = L_1 r^2 \left( 1 + \delta g_{yy} \right) \ , \\
&& g_{rr} = L_1 r^{-2} \left( 1 + \delta g_{rr}\right) \ ,
\end{eqnarray}
where
\begin{eqnarray}
&& \delta g_{tt} = Q_m^2 \left[ \frac{\alpha_t^{(1)}}{r^4} + \frac{\alpha_t^{(2)}}{r^4} \log\left( r \right) \right] \ , \quad \delta g_{xx} = Q_m^2 \left[ \frac{\alpha_x^{(1)}}{r^4} + \frac{\alpha_x^{(2)}}{r^4} \log\left( r \right) \right] \ , \\
&& \delta g_{yy} = Q_m^2 \left[ \frac{\alpha_y^{(1)}}{r^4} + \frac{\alpha_y^{(2)}}{r^4} \log\left( r \right) \right] \ , \quad \delta g_{rr} = Q_m^2 \left[ \frac{\alpha_r^{(1)}}{r^4} + \frac{\alpha_r^{(2)}}{r^4} \log\left( r \right) \right] \ , \\
\end{eqnarray}
with the following constraints:
\begin{eqnarray}
&& \alpha_x^{(2)} = \alpha_t^{(2)} \ , \quad \alpha_y^{(2)} = \alpha_t^{(2)} + \frac{\kappa^2\tau}{2 L_1} \ , \quad \alpha_r^{(2)} = - 4 \alpha_t^{(2)} - \frac{\kappa^2\tau}{L_1} \ , \\
&& \alpha_r^{(1)} = - \alpha_t^{(1)} + \alpha_t^{(2)} - \alpha_x^{(1)} - 2 \alpha_y^{(1)} + \frac{\kappa^2\tau}{3 L_1} \ .
\end{eqnarray}
Since the perturbations grow towards the IR, expectedly the magnetic deformation is a relevant deformation with mass dimension $2$\cite{Kundu:2016oxg}. The logarithmic term directly encodes the information of a conformal anomaly, which is sourced by turning on the magnetic field. In this case, because of the breaking of the Lorentz symmetry at the boundary: SO$(1,3) \to {\rm SO}(1,1) \times {\rm SO}(2)$, the corresponding equation of state will involve energy and two distinct pressures which are parallel and perpendicular to the magnetic field, respectively.

On the other hand, the bulk electric field perturbation yields: 
\begin{eqnarray}
&& \delta g_{tt} = \frac{\varepsilon}{r^4} + Q_e^2  \frac{\beta_t^{(1)}}{r^6} \ , \quad \delta g_{xx} = \frac{p_x}{r^4} + Q_e^2  \frac{\beta_x^{(1)}}{r^6} =  \delta g_{yy} \ , \quad \delta g_{rr} = Q_e^2 \frac{\beta_r^{(1)}}{r^6} \ , \\
&& {\rm and} \quad \partial_r A_t(r) = \frac{Q_e}{\sqrt{L_1}} \frac{1}{r^3} \ ,
\end{eqnarray}
with the following constraints:
\begin{eqnarray}
&& \beta_x^{(1)} = \beta_t^{(1)} - \frac{\kappa^2\tau}{6 L_1^2} = \beta_y^{(1)}  \ , \quad \beta_r^{(1)} = - 6 \beta_t^{(1)} + \frac{5 \kappa^2\tau}{6 L_1^2} \ . 
\end{eqnarray}
In the above, we have included the energy and pressure terms, as well. The bulk gauge field corresponds to turning on a relevant operator of mass dimension $3$ in the CFT. The above was also discussed in details in \cite{Kundu:2016oxg}.

To consider the field, denoted by $P(r)$, note that such a bulk field should correspond to a boundary current on general grounds. The corresponding deformation should have mass dimension $3$, similar to the density perturbation in a $(3+1)$-dimensional CFT. Thus, we can already guess the form of the linear corrections induced by the field $P(r)$, which can subsequently confirmed by an explicit calculation. The result of the linearized analysis can therefore be summarized as below.

The metric perturbations are of the form:
\begin{eqnarray}
&& g_{{tt}} = L_{1}r^{2} \left( 1+ \delta g_{{tt}} \right) \ , \quad  g_{{rr}} = \dfrac{L_{1}}{r^{2}}\left( 1+ \delta g_{{rr}}\right) \ , \\ 
&& g_{yy} = L_{1}r^{2} \left( 1+ \delta g_{yy}  \right) \ , \quad g_{xx} = L_{1}r^{2}\left( 1+ \delta g_{xx} \right)  \ , \\
&& g_{{tx}} = L_{1}r^{2}\delta g_{{tx}} \ , \quad \delta F=  \delta E dr \wedge dt + \delta B dy^1 \wedge dy^2 + \delta P dx \wedge dr \ .
\end{eqnarray}
The solutions are:
\begin{eqnarray}
\delta E = \dfrac{Q_{e}}{r^{3}} \ , \quad  \delta P = \dfrac{Q_{p}}{r^{3}} \ ,
\end{eqnarray}
and 
\begin{eqnarray}
\delta g_{{tt}} & = &  \alpha_{t}r^{-4}+ \beta_{t}r^{-6} + \gamma_{t} r^{-4}~\log r \ , \\
\delta g_{{rr}} & = &  \alpha_{r}r^{-4} + \beta_{ r} r^{-6} + \gamma_{r}r^{-4}~\log r  \ , \\
\delta g_{yy} & = & \alpha_{y} r^{-4} + \beta_{y}r^{-6} + \gamma_{y} r^{-4}~\log r \ , \\
\delta g_{xx} & = & \alpha_{x} r^{-4} + \beta_{x}r^{-6} + \gamma_{x} r^{-4}~\log  r \ , \\
\delta g_{{tx}} & = & \Gamma_1 - \frac{\Gamma_2} {4r^{4}} + \frac{\kappa^{2} \tau ~Q_{e} Q_{p}}{6r^{6}L_{1}} \ , 
\end{eqnarray}
where $\Gamma_{1,2}$ are constants and 
\begin{eqnarray}
\gamma_{r} &=& - 4 \gamma_{t} - \dfrac{\delta B^{2}\kappa^{2}\tau}{L_{1}} \ , \quad  \gamma_{x} = \gamma_{y} - \dfrac{Q_{p}^{2}\kappa^{2}\tau}{6L_{1}} \ , \\
\beta_{t} & = & \gamma_{y} + \dfrac{Q_{e}^{2}\kappa^{2}\tau}{6L_{1}} \ , \quad \beta_{r} = \dfrac{-36 \gamma_{y} L_{1} + \kappa^{2} \tau \left( Q_{p}^{2} - Q_{e}^{2}\right) }{6L_{1}} \ , \\
\beta_{x} & = & \gamma_{t} + \dfrac{\delta B^{2}\kappa^{2}\tau}{2L_{1}} \ , \quad \beta_{x} = \gamma_{t} \ , \\
\gamma_{t} & = & \dfrac{- \delta B^{2}\kappa^{2}\tau + 3 \alpha_{r}L_{1} +  3 \alpha_{t} L_{1} +  6 \alpha_{y} L_{1} +  3 \alpha_{x} L_{1}}{3L_{1}} \ . 
\end{eqnarray}
Thus, one has three independent relevant deformations of the UV CFT, of mass dimension $2$ and $3$ respectively, which are characterized by $7$ free parameters.

The log-terms above correspond to the breaking of conformal symmetry due to the presence of the fluxes. Note that, asymptotically, the magnetic perturbation is dominant over the density (or, current) perturbation simply on the basis of the dimension of the corresponding operator. Furthermore, they also define length scales where various flux deformations become important. For example, the magnetic perturbation becomes important at 
\begin{eqnarray}
{\cal O} \left( 1 \right) = \frac{\delta B^2 \kappa^2 \tau}{L_1} r_{\rm magnet}^{-4} \log r_{\rm magnet} \ , 
\end{eqnarray}
which defines a corresponding energy-scale in the boundary theory. Similarly, the density perturbation becomes important at:
\begin{eqnarray}
{\cal O} \left( 1 \right) = \frac{\delta Q_e^2 \kappa^2 \tau}{L_1} r_{\rm density}^{-6}  \quad \implies \quad r_{\rm density} \sim \delta Q_e^{1/3} \left( \frac{\kappa^2 \tau}{L_1} \right)^{1/6} \ .
\end{eqnarray}
A similar scale can be obtained for the $\delta P$ perturbation. On the other hand, the density perturbation and the magnetic perturbation become of the same order at:
\begin{eqnarray}
{\cal O} \left( \frac{\delta Q_e^2 \kappa^2 \tau}{L_1} r_{\rm density}^{-6} \right) = \frac{\delta B^2 \kappa^2 \tau}{L_1} r_{\rm magnet}^{-4} \log r_{\rm magnet} \ .
\end{eqnarray}
The cross-over scale can be obtained by the principal solutions for $w$ in equation $z = w e^w$, where $z$ is a constant determined by the parameters of our system. This scale sets a hierarchy between the magnetic and the density perturbations.

\section{Appendix 2: Perturbing the AdS$_2$}  \label{appen2}

Let us now carry out the same exercise, around the AdS$_2 \times R^3$ geometry. The linearized perturbation data, as before, is given by
\begin{eqnarray}
&& g_{{tt}} = L_{1}r^{2}\left( 1+  \delta g_{{tt}} \right) \ ,  \quad g_{{rr}} = \dfrac{L_{1}}{r^{2}} \left( 1+  \delta g_{{rr}} \right)  \ , \\  
&& g_{yy} = L_{2} \left( 1+  \delta g_{yy} \right) \ , \quad  g_{xx} = L_{2} \left( 1+  \delta g_{xx} \right) \ ,  \\
&& g_{tx} = L_{1}r ^{2} \delta g_{tx}  \ , \\ 
&& \delta F = \left( Q_e   +  ~\delta Q_e \right) dr \wedge dt  +   \delta B dy^1 \wedge dy^2  + \delta P dx \wedge dr   \ .
\end{eqnarray}
Solving the zeroth order field equations, we obtain:
\begin{eqnarray}
E = Q_e \ , \quad \Lambda &=& -\dfrac{L_{1}}{Q_e^{2}} \ , \quad  \kappa ^{2} \tau = \dfrac{\sqrt{L_{1}^{2}-Q_e^{2}}}{Q_e^{2}} \ . 
\end{eqnarray}

Now, the solution of the linearized gauge field equations yield:
\begin{eqnarray}
\delta E & = &  \delta Q_{e} \hspace{2mm}({\rm which} \hspace{2mm} {\rm is} \hspace{2mm} {\rm arbitrary} ) \ , \\
\delta P &=& \frac{\delta Q_p}{r^3}  \ .
\end{eqnarray}
The metric perturbations are:
\begin{eqnarray}
&& \delta g_{{tt}} = \alpha_{t} r^m  \ , \quad \delta g_{{rr}} = \alpha_{r} r^m \ , \quad \delta g_{yy} = \alpha_{y} r^m \ , \\
&& \delta g_{xx} = \alpha_{x} r^m \ , \\
&& \delta g_{tx} = \frac{C_1}{r^2} + \frac{\delta Q_p}{Q_e} \frac{1}{r^3} - \frac{4 \alpha  \delta B \sqrt{L_1^2- Q_e^2}}{L_1^2 L_2^2 r} \ , \\
\end{eqnarray}
where $C_1$ is a free constant that can be independently set to zero. Now, there are various classes of solutions. These are given by
\begin{eqnarray}
\alpha_t = 2 \frac{\delta Q_{e}}{Q_e} \ , \quad \alpha_r = 0 \ , \quad m= 0 \ . \label{ads2pert1}
\end{eqnarray}
\begin{eqnarray}
\alpha_t = 2 \frac{\delta Q_{e}}{Q_e(m+1)} \ , \quad \alpha_r = \frac{2 m \delta Q_e}{(m+1) q} \ , \quad \alpha_x = 0 \ , \quad \alpha_y = 0 \ . \label{ads2pert2}
\end{eqnarray}
\begin{eqnarray}
\alpha_r = \frac{2 \delta Q_{e} - \alpha_t Q_e}{Q_e} \ , \quad \alpha_x = 0 = \alpha_y  \ , \quad m = -2 \ .  \label{ads2pert3}
\end{eqnarray}
\begin{eqnarray}
\alpha_t & = & \frac{\alpha_x L_1^2 Q_e - 3 \alpha_x Q_e^3 + 3 L_1^2 \delta Q_{e}}{3 L_1^2 Q_e} \ , \label{ads2pert41} \\
\alpha_r & = & \frac{8 \alpha_x L_1^2 Q_e - 6 \alpha_x Q_e^3 + 3 L_1^2 \delta Q_{e}}{3 L_1^2 Q_e} \ , \label{ads2pert42} \\
\alpha_x & = & \alpha_y \ , \quad m = 1 \ . \label{ads2pert43}
\end{eqnarray}
The solution in (\ref{ads2pert3}), in particular, corresponds to the introduction of an event horizon in the geometry. It is also clear that in (\ref{ads2pert41})-(\ref{ads2pert43}) both relevant and irrelevant modes are turned on, similar to \cite{Kundu:2016oxg}\footnote{Note, however, that the perturbation series carried out in \cite{Kundu:2016oxg} is not precisely the same as here.}, and can therefore naturally define two length-scales where the back-reaction can become order unity. The irrelevant modes do not decouple from the linearized analysis, which essentially signifies that AdS$_2$ asymptotic structure cannot be supported for any state other than the ground state. Note further that the relevant perturbation in $\delta g_{tx}$ can not be decoupled from the IR dynamics, even if we choose to set $\delta Q_p = 0 = \delta Q_e$, due to the non-vanishing Chern-Simons coefficient. In fact, in the presence of both $\delta Q_p$ and $\delta B$, one observes a competition between the two deformations and the possibility of a non-trivial IR configuration, perhaps outside the scope of an analytical handle.

\section{Appendix 3: Perturbing the AdS$_3$}  \label{appen3}

The full perturbed metric and flux, as before, reads:
\begin{eqnarray}
&& g_{{tt}} = L_{1}r^{2} \left( 1+ \delta g_{{tt}} \right) \ , \quad  g_{{rr}} = \frac{L_1}{r^{2}}\left( 1+ \delta g_{{rr}} \right) \ , \\  
&& g_{yy} = L_{2}\left( 1+  \delta g_{yy} \right) \ , \quad g_{xx} = L_{1}r^{2}\left( 1+  \delta g_{xx} \right) \ , \\
&& g_{tx}(r) = L_{1}r^{2} \delta g_{tx} \ , \quad \delta F=  ~ \delta E dr \wedge dt  + \delta P dx \wedge dr \ . 
\end{eqnarray}
The zeroth order solution is given by
\begin{eqnarray}
\Lambda  =  -\frac{\frac{2 L_2^2}{B^2}+3}{L_1} \ , \quad  \kappa ^{2} \tau = \frac{2 L_2 \sqrt{B^2+L_2^2}}{B^2 L_1} \ . 
\end{eqnarray}
The gauge field perturbations can be solved to obtain:
\begin{eqnarray} \label{ads3gaugepert}
\delta E  = \frac{\delta Q_{e} e^{-\frac{\sqrt{3} \alpha }{r}}}{2 r} = \delta P \ , \quad {\rm with} \quad L_1 = 2 \ , \quad L_2 = 1 \ , \quad B = \sqrt{3} \ .
\end{eqnarray}
Solving the linearized Einstein equations, one obtains:
\begin{eqnarray}
\delta g_{tt} & = & \frac{\alpha_t}{r^2} + \frac{\beta_t}{r^3} + \frac{\gamma_t \log (r)}{r^2} \ , \\
\delta g_{rr} & = & \frac{\alpha_r}{r^2} + \frac{\beta_r}{r^3} + \frac{\gamma_r \log (r)}{r^2} \ , \\
\delta g_{yy} & = & \frac{\alpha_y}{r^2} + \frac{\beta_y }{r^3} + \frac{\gamma_y \log (r)}{r^2} \ , \\
\delta g_{xx} & = & \frac{\alpha_x}{r^2} + \frac{\beta_x}{r^3} + \frac{\gamma_x \log (r)}{r^2} \ . 
\end{eqnarray}
The various constants are given by
\begin{eqnarray}
&& \alpha_y = 0 = \beta_y = \gamma_y \ , \quad \beta_x = \frac{2 \alpha  \delta Q_{e}^2}{3 \sqrt{3}} - \frac{\beta_r}{3} \ , \quad \beta_t = - \frac{\beta_r}{3} - \frac{2 \alpha  \delta Q_{e}^2}{3 \sqrt{3}} \ , \\
&&  \gamma_x = \frac{1}{6} \left(6 \alpha_r + 6 \alpha_t + 6 \alpha_x + \delta Q_{e}^2\right) \ , \quad \gamma_t = \frac{1}{6} \left(6 \alpha_r + 6 \alpha_t + 6 \alpha_x - \delta Q_{e}^2\right) \ , \\
&& \gamma_r = - \gamma_t - \gamma_x \ .
\end{eqnarray}
Note that, the perturbations in (\ref{ads3gaugepert}) take a form similar to \cite{Donos:2014gya}, in which the dimension of the corresponding operator is readable only when the exponential is expanded in the infinite series, and therefore corresponds to turning on an infinite number of relevant deformations.

The linearized fluctuations can be parametrized by the following data: $\left\{ \alpha, \beta, \gamma \right\}$. Similar to what was observed in \cite{Kundu:2016oxg}, viewed from the local CFT$_2$ perspective, the metric deformation corresponds to relevant operators of mass dimension $2$ and a new relevant deformation of mass dimension $3$. The dimension $2$ deformation corresponds to the boundary stress-tensor, and the dimension $3$ deformation is sourced by the boundary current/density, denoted by $\delta P = \delta Q_e$. As before, all these deformations grow towards the IR, and subsequently define a natural scale associated with each of them, obtained by setting the linear order correction to be order unity. We can certainly kill of all the $\alpha$'s, some of the $\beta$'s and $\gamma$'s in the coefficients of the various perturbation modes. The presence of the log-term, along with the power law, naturally defines two energy-scales: one where the density perturbation becomes important around the CFT$_2$, and one which encodes the breaking of conformal symmetry sourced by the flux.

\section{Appendix 4: Perturbing warped ${\rm AdS}_{3}$} \label{appen4}

The linearized metric and flux data are, as before, given by
\begin{eqnarray}
&& g_{{tt}} = \left(L_{1} - L_{2}^{2} \right) r^{2}\left(1+\delta g_{tt} \right)  \ , \quad g_{{rr}} = \frac{1}{L_{1}r^{2}}\left(1+ \delta g_{rr}  \right) \\ 
&& g_{yy} = \left( 1+ \delta g_{yy} \right) \ , \quad g_{xx} =\left( 1+ \delta g_{xx} \right) \ , \quad g_{tx} = L_{2}r\left(1+ \delta g_{tx}  \right) \ , \\ 
&& \delta F=  \left( E+ \delta E \right) dr \wedge dt  +  B ~ dy^1 \wedge dy^2  + \delta P ~ dx \wedge dr  \ . 
\end{eqnarray}
The zeroth order solution is described by
\begin{eqnarray}
L_{1} &=& -\frac{\left( \kappa^2 \tau \right) ^2}{8 \alpha^2 } - 8 \alpha^2 - 2 \Lambda \ , \quad L_{2} = 2 \sqrt{-8 \alpha^2 -\Lambda } \ , \\
 q & = & \dfrac{\sqrt{1+B^{2}}L_{2}}{4B\alpha} \ , \quad B = \sqrt{-\frac{\alpha^2 (8 \alpha^2 + \Lambda )}{\left( \kappa^2 \tau \right )^{2} }} \ . 
\end{eqnarray}
For simplicity, we choose: $B=1$, $q=\frac{1}{2}$, without the loss of any generality.

One can consider a truncation to zero flux perturbation: $\delta P = 0 = \delta E$. In this case, the metric perturbations are obtained by
\begin{eqnarray}
&& \delta g_{tt} = \alpha_t r^m \ , \quad \delta g_{yy} = \alpha_y r^m \ , \quad g_{xx} = \alpha_x r^m \ , \\
&& \delta g_{rr} = \alpha_r r^m \ , \quad \delta g_{tx} = \alpha_{tx} r^m \ . 
\end{eqnarray}
The solutions are:
\begin{eqnarray}
m = - 2 \ , \quad \alpha_t = - 4 \alpha_r \ , \quad \alpha_x = 0 = \alpha_y = \alpha_{tx} \ .
\end{eqnarray}
The above perturbation corresponds to event horizons, as we have explicitly constructed in the main text as well.

Considering the non-vanishing flux perturbations, we get the following solution:
\begin{eqnarray}
\delta P = \delta Q_p \,  r^{m} \ , \quad \delta E = \delta Q_e \,  r^{m+1} \ ,
\end{eqnarray}
\begin{eqnarray}
&& \delta g_{tt} = \alpha_t r^{m+1} \ , \quad \delta g_{yy} = \alpha_y r^{m+1} \ , \quad g_{xx} = \alpha_x r^{m+1} \ , \\
&& \delta g_{rr} = \alpha_r r^{m+1} \ , \quad \delta g_{tx} = \alpha_{tx} r^{m+1} \ ,
\end{eqnarray}
where 
\begin{eqnarray}
&& \delta Q_e = - \frac{\sqrt{\frac{2}{3}} \alpha  ( m (4 m (8 m (4 m + 21) + 263) + 459) + 20)}{32 m (m + 3) + 35} \ , \\
&& \alpha_t = - \frac{8 \sqrt{\frac{2}{3}} \alpha  ( m (m (2 m (32 m ( m + 9) + 905) + 2315) + 1015) + 73)}{ (m + 2) (m + 3) (32 m (m + 3) + 35)} \ , \\
&& \alpha_y = - \frac{52 \sqrt{\frac{2}{3}} \alpha }{32 m (m + 3) + 35} \ , \quad \alpha_x = \frac{104 \sqrt{\frac{2}{3}} \alpha }{32 m (m + 3) + 35} \ , \\
&& \alpha_{tx} = - \frac{8 \sqrt{\frac{2}{3}} \alpha  ( m ( m (32 m ( m + 6) + 393) + 302) + 53)}{( m + 2) (32 m ( m + 3) + 35)} \ , \\
&& \alpha_r = -\frac{2 \sqrt{\frac{2}{3}} \alpha  (4 m ( m (2 m ( m (32 m (m + 10) + 1241) + 2351) + 4481) + 1948) + 1067)}{( m + 2) ( m + 3) (32 m (m + 3) + 35)} \ , \nonumber\\
\end{eqnarray}
where $m$ remains undetermined by the equations of motion.

If we want growing modes in the metric perturbations, the flux perturbation (specifically, $\delta E$) also grows. On the other hand, for suitably chosen negative values of $m$, the perturbations above, both flux and the metric, grows in the IR.

\section{Appendix 5: Schrodinger Solutions} \label{appen5}

Before perturbing the Schr\"{o}dinger geometry, we will fix the parameters such that our solutions resemble the one discussed in \cite{D'Hoker:2009bc}. This is achieved by setting: $B=\sqrt{3}$, $L_1 = 2 B = 2\sqrt{3}$, $L_2 = 1$, $\Lambda = - 11$, $\kappa^2 \tau = 4$. The perturbed geometric data are given by
\begin{eqnarray}
g_{{tt}} &= & \dfrac{8r^{-2\alpha }q^2}{\alpha + 2\alpha^{2}} + \delta g_{tt}  \ , \quad g_{{rr}} =  \dfrac{1}{12r^{2}} + \delta g_{{rr}}  \ , \\
g_{yy} &= & \left( 1+ \delta g_{yy} \right) \ , \quad g_{xx} =  \left( 1+ \delta g_{xx} \right) \ , \\
g_{tx} &= & 2\sqrt{3}r + \delta g_{tx} \ , \quad F= \left( E+ \delta E \right) dr \wedge dt  +   B  dy^1\wedge dy^2  +  \delta P dx \wedge dr \ ,
\end{eqnarray}
where $E = q r^{-\alpha -1}$ .

The solution for the linearized fluxes is give by
\begin{eqnarray}
&& \delta P = \delta Q_{p} ~ r^{\alpha - 1 } \ , \quad  \delta E   =  \delta Q_e \, r^{n - 1 - \alpha} \ .
\end{eqnarray}
The metric perturbations are given by
\begin{eqnarray}
\delta g_{tt} &= &  \alpha_t r^n \ ,  \quad \delta g_{yy} =  \alpha_y r^n  \ , \quad \delta g_{rr}  =  \alpha_r r^n  \ , \\
\delta g_{tx} & = & \alpha_{tx} r^n  \ , \quad \delta g_{xx}  =  0  \ .
\end{eqnarray}
The algebraic equations involving the various constants can be solved by
\begin{eqnarray}
&& n = -1 \ , \quad \alpha_y = \frac{40 Q_e \, \delta Q_p}{13 \sqrt{3}} \ , \\
&& \alpha_t = \frac{2 \left(39 \alpha  \delta Q_e + \frac{4 \sqrt{3} (\alpha  (12 \alpha -1)-10) Q_e^2 \, \delta Q_p}{2 \alpha +1}\right)}{39 (\alpha +1) Q_e} \ , \\
&& \alpha_r = \frac{78 \alpha  (2 \alpha +1) \delta Q_e + 8 \sqrt{3} (\alpha  (8 \alpha  (7 \alpha + 9) + 9)-13) Q_e^2 \, \delta Q_p}{39 \alpha  (\alpha +1) (2 \alpha +1) Q_e}  \ , \\ 
&& \alpha_{tx} = - \frac{39 \alpha  (2 \alpha +1) \delta Q_e + 4 \sqrt{3} \left(\alpha  \left(64 \alpha ^2 + 84 \alpha + 13\right) - 13\right) Q_e^2 \, \delta Q_p}{39 \alpha  (\alpha +1) (2 \alpha +1) Q_e} \ .
\end{eqnarray}
We can freely set $\delta g_{xx} = 0$, which we have done above. However, one can obtain a general solution:
\begin{eqnarray}
\delta g_{xx} = C_1 + C_2 r \ , 
\end{eqnarray}
which results in a more general solution for $\delta P$:
\begin{eqnarray}
\delta P = \delta Q_p r^{\alpha - 1} + d_1 C_1 r^{- \alpha - 1 } + d_2 C_2 r^{-\alpha - 2} \ . 
\end{eqnarray}
In the above $C_{1,2}$ and $d_{1,2}$ are independent constants. However, we are free to set $C_{1,2} = 0 $. Metric perturbations have growing modes in the IR: it is unlikely that this growing mode corresponds to turning on a non-vanishing temperature, since these perturbations are supported by flux perturbations only. On the other hand, $\delta g_{xx}$ seems to gave a growing mode, independent of the flux perturbations.

\end{document}